\documentclass[%
 aip,
 amsmath,amssymb,
 reprint,%
]{revtex4-1}

\usepackage{graphicx}
\usepackage{dcolumn}
\usepackage{bm}
\usepackage{upgreek}
\usepackage[utf8]{inputenc}
\usepackage[T1]{fontenc}
\usepackage{mathptmx}
\usepackage{etoolbox}
\usepackage{siunitx}
\usepackage{soul}

\makeatletter
\def\@email#1#2{%
 \endgroup
 \patchcmd{\titleblock@produce}
  {\frontmatter@RRAPformat}
  {\frontmatter@RRAPformat{\produce@RRAP{*#1\href{mailto:#2}{#2}}}\frontmatter@RRAPformat}
  {}{}
}%
\makeatother
\begin{document}

\preprint{AIP/123-QED}

\title[Simulation of surface x-ray emission from the ASTERICS ECR ion source]{Simulation of surface x-ray emission from the ASTERICS ECR ion source}
\author{T. Thuillier}
\email{thomas.thuillier@lpsc.in2p3.fr.}
 \author{A. Cernuschi}
 \author{B. Cheymol}%
\affiliation{Université Grenoble Alpes, CNRS-LPSC, INP Grenoble, Grenoble, France
}%

\date{\today}

\begin{abstract}
The bremsstrahlung x-ray emission induced by the impact of plasma electrons de-confined on the chamber wall of the ASTERICS electron cyclotron resonance ion source is investigated through a suite of two simulation codes. The electron high energy temperature distribution tail at the wall is found to be anisotropic and increases with $B_{min}$. The electrons impinge the walls with broad angular distribution peaking at angles ranging between 5-25° with respect to the surface, which has consequences on the x-ray emission directionality and on the yield of electrons bouncing back toward the plasma, reaching up to 50\%.  The x-ray dose is mapped inside and around the ion source for $B_{min}$\,=\,\SI{0.8}{\tesla} and an electron temperature artificially increased to \SI{120}{\kilo\electronvolt} to dimension with margin the cave shielding. The dose without shielding  reaches  \SI{100}{\micro\sievert\per\hour} per kW of impacting electrons at 5 m. A set of internal and external shielding is presented to attenuate this dose and reduce it to less than \SI{1}{\micro\sievert\per\hour} per kW of electrons. A parametric electron distribution temperature study with Fluka indicates that the deposition of 1 W of heat in the superconducting cold mass per kW of plasma electrons, as reported experimentally, is obtained when the temperature is set to \SI{380}{\kilo\electronvolt}. Such a result is compatible with previous experiments achieved on several ion sources showing an x-ray spectral temperature 3 to 4 times higher radially.

\end{abstract}

\maketitle

\section{\label{sec:level1}Electron losses to the wall}
\subsection{Model and methodology}
\begin{figure}[!ht]
\centering
\includegraphics[width=0.48\textwidth]{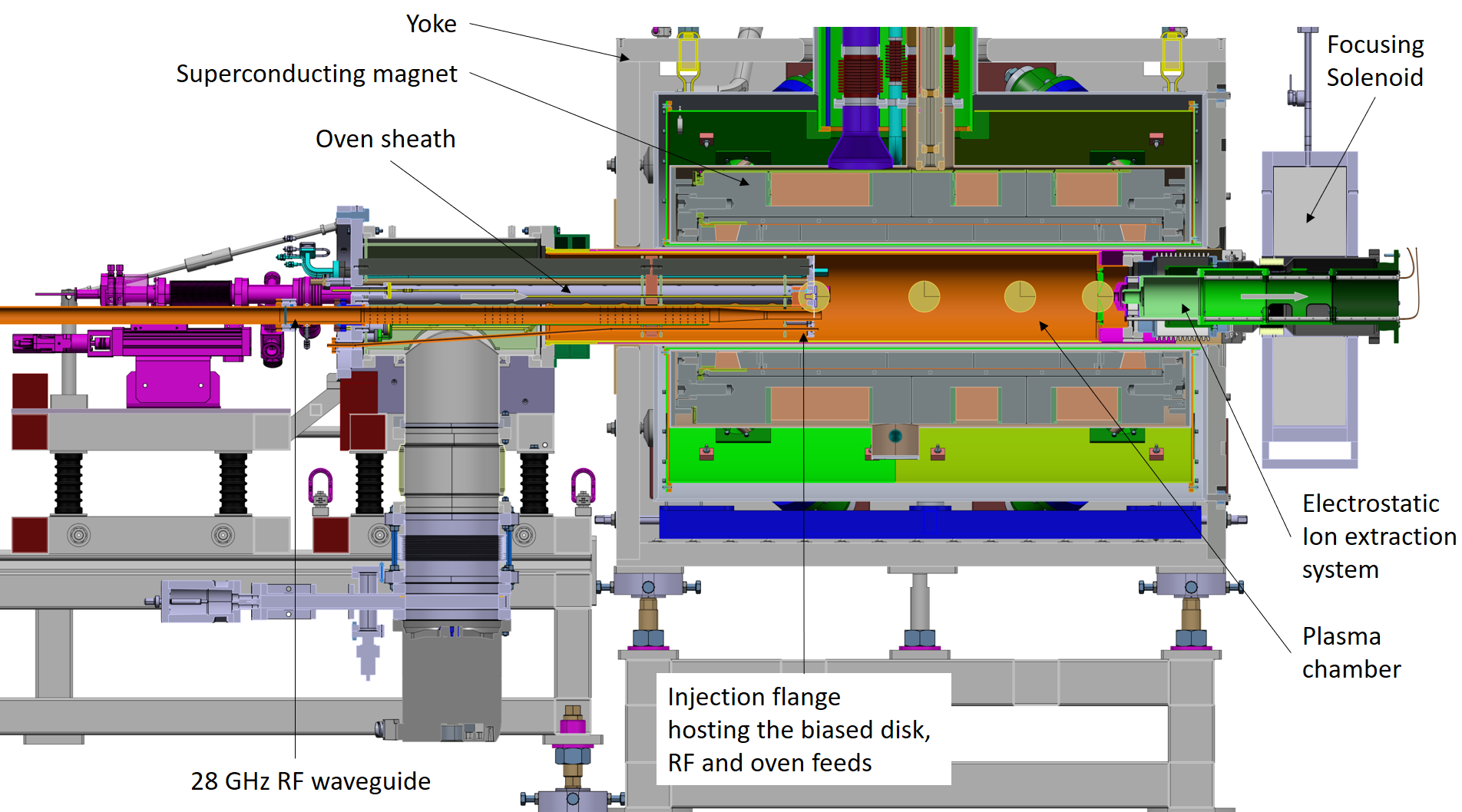}
\caption{Cutaway view of the ASTERICS ion source design.}
\label{fig:FIG1}
\end{figure}

An existing Monte-Carlo (MC) code was adapted to study the electron dynamics inside the ASTERICS ion source plasma chamber~\cite{bib:thuillier_lbl,bib:lpsc_asterics}, currently under design (see the overall source cutaway view on Fig.~\ref{fig:FIG1}). The ASTERICS ion source will be fed with a 28 GHz radio-frequency (RF) circularly polarized electromagnetic wave which is modeled in the simulation with a transverse traveling plane wave with a circular polarization and a constant electric field intensity $E$\,=\,\SI{10}{\kilo\volt\per\meter} (corresponding to 7~kW of injected RF power). An argon plasma is assumed with a mean ion charge state of $\sim$8. The plasma density considered is 15\,\% of the cut-off density at 28 GHz. The ions are considered frozen in the cavity. 
The initial electron energies are randomly sampled using a set of Gaussian distributions centered on each argon ionization potential energies (IP) with a standard deviation of $10\,\%\, \times\,$IP, with a relative abundance following a typical argon ion spectrum. The electrons' initial velocity direction is randomly and isotropically distributed is space. 
The electrons undergo Coulomb collision and electron impact with ions. The RF pitch angle~\cite{bib:Cluggish} scattering is not considered in the simulation, in order to maximize the electrons energy and the associated high energy bremsstrahlung emission. The electrons are tracked until they impact the 3 possible walls: injection at $z_{inj}$\,=\,\SI{0.3}{\meter}, extraction at $z_{ext}$ \,=\,\SI{-0.3}{\meter} and radial wall at $r_W$\,=\,\SI{0.091}{\meter}. Two static electric fields are modeled in the MC simulation. One for the injection biased disk with a voltage of -100 V and a diameter of 20 mm. The second for the accelerating electric field of the ion source on the extraction, being \SI{10}{\kilo\volt\per\centi\meter}, extending for \SI{4}{\centi\meter} right after the extraction electrode hole of \SI{10}{\milli\meter} diameter. The electrons having a propagation time larger than \SI{1}{\milli\second} (this time being 3-4 times larger than the mean electrons confinement time simulated) are considered highly confined and are stopped. A set of $1.25\times10^6$ electrons was simulated for each magnetic configuration. The electron final positions and velocities on the plasma chamber wall were stored and analyzed.
The electron particle distribution at the plasma chamber wall was studied for two  axial magnetic field configurations: 3.7-0.3-2.2 T and 3.7-0.8-2.2 T, deemed representative of the actual ion source operation. The hexapolar radial magnetic field intensity at wall is fixed at \SI{2.4}{\tesla}. The $B_{min}$\,=\,\SI{0.3}{\tesla} configuration, suitable for double frequency operation (18+28 GHz), is known to generate a low output flux of energetic x-rays~\cite{bib:lbl_bmin}. 
On the contrary, the $B_{min}$\,=\,\SI{0.8}{\tesla} configuration experimentally maximises the production of high energy x-rays~\cite{bib:lbl_bmin,bib:lbl_xray_msu,bib:lbl_xray_imp}. The aforementioned magnetic configurations are used to probe the minimum and maximum x-ray dose in the accelerator cave, respectively. 
\subsection{Monte-Carlo Results}
\begin{figure}[!ht]
\centering
\includegraphics[width=0.48\textwidth]{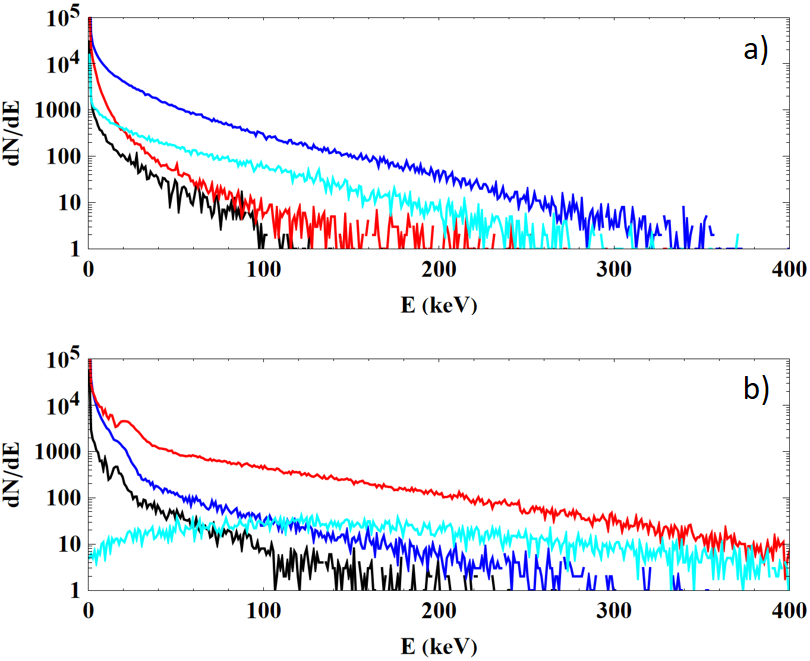}
\caption{EEDF of the electrons hitting the plasma chamber wall for \textbf{(a)} $B_{min}$\,=\,0.3~T and \textbf{(b)} $B_{min}$\,=\,0.8~T. The black, blue and red plots are respectively recorded on the injection (z=$z_{inj}$), radial ( r=$r_{wall}$) and extraction surfaces (z=$z_{ext}$). The cyan EEDF plots corresponds to the electron still confined in the plasma chamber volume after 1 ms.}
\label{fig:FIG2}
\end{figure}

\begin{figure}[!ht]
\centering
\includegraphics[width=0.48\textwidth]{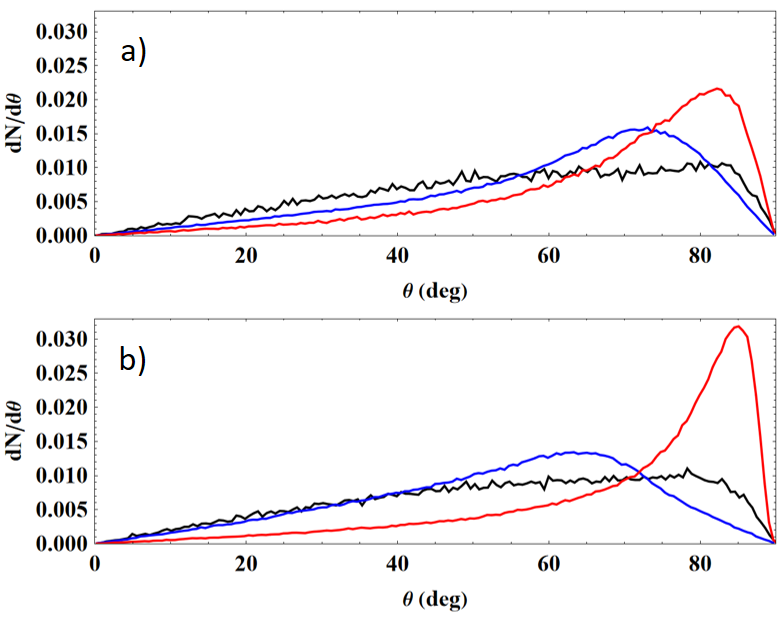}
\caption{Distribution of the angle of incidence of electron impacting the 
plasma chamber walls ($\theta=(\widehat{\vec{v},\vec{n}}$)), $\vec{n}$ normal to the
surface) \textbf{(a)} for $B_{min}$\,=\,\SI{0.3}{\tesla} and  \textbf{(b)} for $B_{min}$\,=\,\SI{0.8}{\tesla}. The black, blue and red curves correspond to the injection, radial and extraction surfaces respectively.}
\label{fig:FIG3}
\end{figure}

\begin{figure*}[!tbh]
    \centering
    \includegraphics*[width=\textwidth]{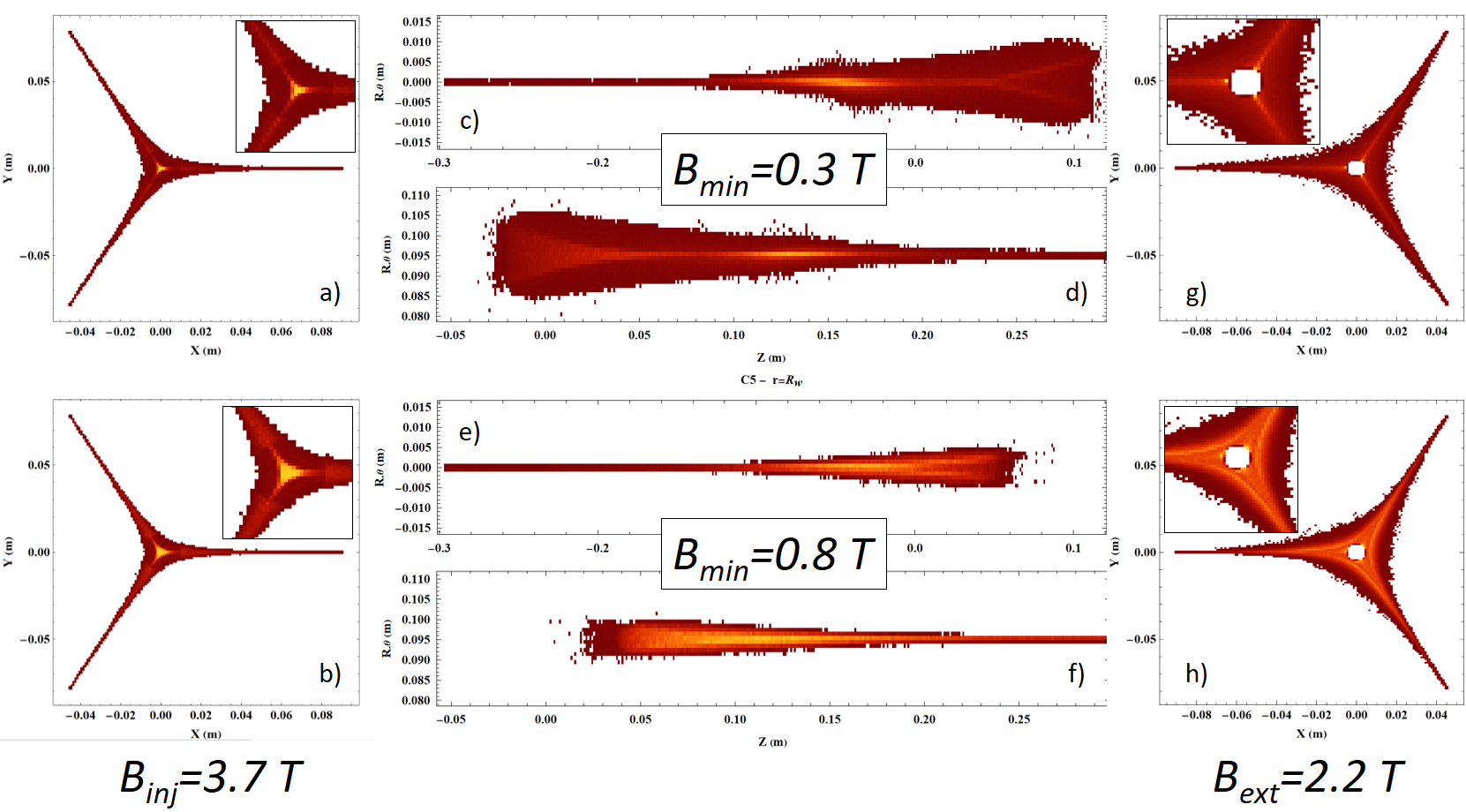}
    \caption{Electron density distribution at the injection ((a) and (b)), radial ((c), (d), (e) and (f)) and extraction ((g) and (h)) surface of the plasma chamber wall for $B_{min}$\,=\,\SI{0.3}{\tesla} (top plots) and $B_{min}$\,=\,\SI{0.8}{\tesla} (bottom plots). The dimension scale of images between the top and the bottom  is conserved.}
    \label{fig:FIG4}
\end{figure*}

The electron energy distribution function (EEDF) of the electrons impacting the walls are plotted in Fig.~\ref{fig:FIG2}(a) and Fig.~\ref{fig:FIG2}(b) for the magnetic configurations $B_{min}$\,=\,0.3 and 0.8~T respectively. The EEDF for the injection, radial and extraction walls are reported in each subplot in black, blue and red respectively. For completeness, the EEDF of electrons still confined in the source volume after 1 ms are indicated in cyan. This specific electron population is expected to contribute to the volume plasma bremsstrahlung emission which is not studied here. The high energy part of each EEDF population have been fitted with a Maxwell-Boltzmann distribution (above 80 to 100 keV, depending on the statistics), and the temperature obtained, expressed in keV, are reported in the Tab.~\ref{tab:temp_surface} for each surface and magnetic configuration and for the electron still confined after 1 ms. The EEDF is found to vary both on the wall surface location and with  $B_{min}$. The normalized counts per wall surface associated with  $B_{min}$\,=\,0.3 and 0.8~T are proposed in Tab.~\ref{tab:distr_position}. One can note a transfer of the flux of electrons from the radial wall (76 to 52\,\%) to the extraction wall (17 to 41\,\%) when $B_{min}$ is changed from 0.3 to 0.8~T. The reason for this shift is a change of the minimum magnetic field intensity at the plasma chamber wall, which passes from 2.03  to 2.29 T, making the weakest magnetic point the extraction peak field (2.2 T) for the latter case. The increase of $B_{min}$ is coming along with a temperature increase of the hot electrons ($T_{rad}$ from 41 to 52 keV, $T_{ext}$ from 44 to 63 keV). This specific topic will be discussed in another paper dedicated to the ASTERICS plasma x-ray volume emission~\cite{bib:cernuschi_24}. It is also worth noting that, for $B_{min}$\,=\,\SI{0.8}{\tesla}, the three EEDF feature a visible hump for E$\approx$15-20 keV, which is known to cause plasma instabilities and has been confirmed experimentally for such a high $B_{min}$~\cite{bib:Tarvainen_instability}. Figure~\ref{fig:FIG3} presents the distribution of angle of incidence $\theta$ of electrons hitting the plasma chamber wall ($\theta=(\widehat{\vec{v},\vec{n}}$),  $\vec{n}$ local normal to the surface) for (a) $B_{min}\,=\,0.3$ T and (b) 0.8~T. The color plot convention in  Fig.~\ref{fig:FIG3} is identical to the one adopted in Fig.~\ref{fig:FIG2}. One can observe how the magnetic field intensity strongly influences the distribution shape. While the distributions for the injection are almost identical for (a) and (b), one can observe a stronger peaking of the extraction wall distribution when $B_{min}$ is increased from 0.3 to 0.8~T, with a most probable angle of $\approx$ 85°. On the other hand, a concomitant reduction of the most probable impact angle at the radial wall is found, from 72° to 65°. The high values of $\theta$ are a consequence of the magnetic mirror effect (pitch angle loss cone) and the concomitant accumulation of perpendicular electron energy by the ECR mechanism. 
The direction of impact of electrons on the wall influences the direction of emission of bremsstrahlung photons and must  be considered in the bremsstrahlung simulation. The electron distribution on the injection, radial and extraction wall is proposed for $B_{min}$\,=\,0.3~and~0.8~T in Fig.~\ref{fig:FIG4}. While the electron distribution is marginally affected on the injection surface (with nevertheless a triangular core centered on the axis which is twice as large as the case when $B_{min}$\,=\,\SI{0.8}{\tesla}), one can observe how the distribution is significantly re-balanced between the radial and the extraction walls. At $B_{min}$\,=\,\SI{0.8}{T}, the radial electron distribution is concentrated on a much smaller surface, while the place where high flux of electron hits the extraction wall is largely enhanced.
\begin{table}[!hbt]
\begin{ruledtabular}
\caption{\label{tab:temp_surface} Estimation of the EEDF high energy tail temperature obtained on the injection ($T_{inj}$), radial ($T_{rad}$) and extraction plasma chamber walls ($T_{ext}$) for the two cases $B_{min}$\,=\,0.3 and 0.8~T. $T_{conf}$ corresponds to the temperature of the electrons still confined after 1 ms.}
\begin{tabular}{lcccr}
Axial profile & T$_{inj}$ & T$_{rad}$ & T$_{ext}$& T$_{conf}$\\
\,\,\,\,\,\,\,\;(T) & (keV) & (keV) & (keV) & (keV)\\
\hline
3.7-0.3-2.2  &  $19.8\pm 1.6$ & $41.7\pm 1.5$ & $44.0\pm 7.4$ &$39.2\pm 1.5$\\
3.7-0.8-2.2  &  $36.0\pm 8.6$ & $52.0\pm 3.1$ & $63.2\pm 7.9$ &$89.2\pm 1.6$\\
\end{tabular}
\end{ruledtabular}
\end{table}
\begin{table}[!hbt]
\caption{\label{tab:distr_position}%
Distribution of the final position of the electrons for the two magnetic axial profiles considered. The subscripts  \%\textit{inj},  \%\textit{ext} and \%\textit{rad} refer respectively to the particles deconfined at the injection, extraction and radial walls.  \%\textit{conf} stands for the amount of electrons still confined at the time limit of 1 ms. The statistical uncertainty on the \% values is $\sim 0.5\%$.}
\begin{ruledtabular}
\begin{tabular}{lcccr}
Axial profile&
\%$_{inj}$&
\%$_{ext}$&
\%$_{rad}$&
\%$_{conf}$\\
\hline
3.7-0.3-2.2 T & 3\% & 17\% & 76\% & 4\%\\
3.7-0.8-2.2 T  & 6\% & 41\% & 52\% & 1\%\\
\end{tabular}
\end{ruledtabular}
\end{table}
\begin{table}[!hbt]
\caption{\label{tab:photonyield}
Total x-ray photon to electron yield passing through the injection, radial and extraction wall surfaces, for $B_{min}$\,=\,0.3 and 0.8~T. The statistical uncertainty on the \% values is $\sim 0.5\%$.}
\begin{ruledtabular}
\begin{tabular}{lcccr}
Axial profile & photon\,/\,$e^{-}$&\%$_{inj}$&\%$_{rad}$ &\%$_{ext}$\\
\hline
3.7-0.3-2.2 T & $5.7\,10^{-5}$ & 15\,\% & 60\,\% & 25\,\%\\
3.7-0.8-2.2 T & $1.6\,10^{-4}$ & 27\,\% & 11\,\% & 62\,\%\\
\end{tabular}
\end{ruledtabular}
\end{table}

\begin{figure*}[!ht]
\centering
\includegraphics[width=\textwidth]{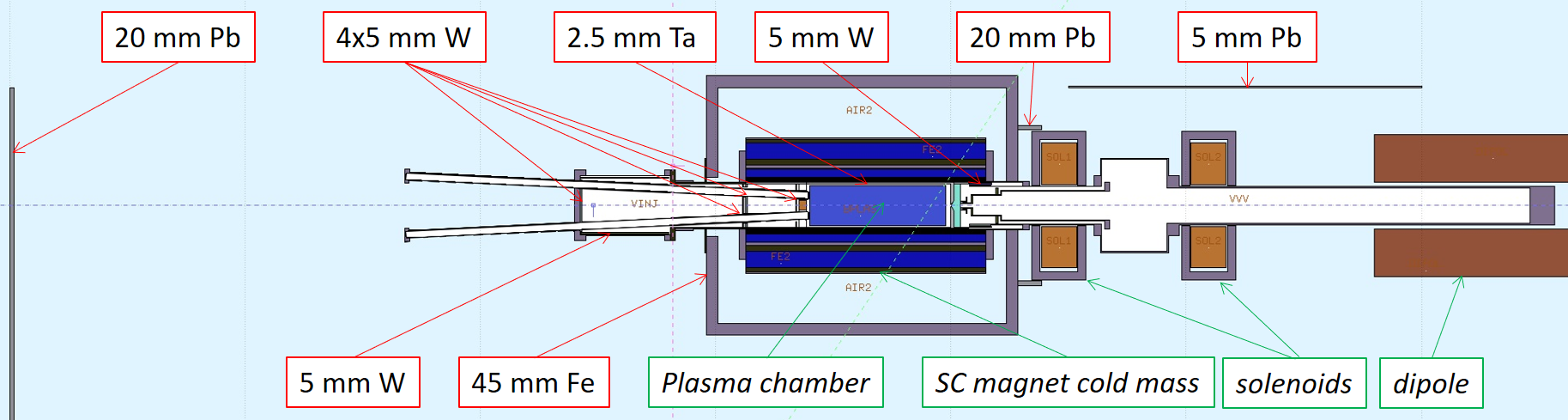}
\caption{Sectional view of the ion source geometry modeled with Fluka. Detail of the materials thickness used to shield the x-ray emission from the source are provided. See text for details.}
\label{fig:FIG5}
\end{figure*}
\section{\label{sec:level2}Fluka Simulation}
\subsection{\label{sec:flukamodel} Geometrical model and methodology}
A simplified  three dimensions version of the ASTERICS ion source geometry and the first meter of its low energy beam transfer line (LEBT) have been modeled with Fluka~\cite{bib:fluka}. A sectional view of the Fluka model is proposed in  Fig.~\ref{fig:FIG5}. The ion source model includes the actual geometry of the superconducting magnet cold mass~\cite{bib:cea_asterics}, the source iron yoke, the plasma chamber, representative components of the atom and RF injection system (two oven ports and one 28 GHz over-sized waveguide) and the ion extraction electrodes. The LEBT model includes the elements located on the ion source axis: the beam pipes, a vacuum chamber, the first two solenoids and a 90° bending magnet. The model also includes the actual ion source cave geometry whose walls are made up with concrete. All the materials and compounds are defined in the simulation according to the actual ones, or by the closest achievable in Fluka database (e.g. pure aluminum instead of 6061 alloy grade). The Fluka model is completed with a set of shields located at specific places to damp the bremsstrahlung emission induced by the impact of electrons on the plasma chamber surface. The shield locations and their thicknesses are indicated in Fig.~\ref{fig:FIG5}. The goal of these added shields is to mitigate the dose around the ion source, allowing the safe circulation of personnel in a corridor located next to the ion source, with a maximum dose rate of \SI{7.5}{\micro\sievert\per\hour}. 
The output of the MC electron simulation (position and velocity direction) is used as an input for the Fluka code, while the EEDF is assumed to follow a Maxwell-Boltzmann distribution with a chosen temperature.
The MC results for $B_{min}$\,=\,\SI{0.3}{\tesla} with an EEDF temperature fixed to \SI{50}{\kilo\electronvolt} are used to study a typical operation which minimizes the x-ray dose emitted from the source. The MC results for $B_{min}$\,=\,\SI{0.8}{\tesla} are used to study the case of high bremsstrahlung x-ray flux in operation and dimension the x-ray shieldings necessary for the safe  ion source operation and in the cave. For $B_{min}$\,=\,\SI{0.8}{\tesla}, the EEDF temperature is set to \SI{120}{\kilo\electronvolt}, a value 20\,\% higher than the measured experimental axial x-ray spectral temperature in similar ion sources~\cite{bib:lbl_bmin,bib:lbl_xray_imp,bib:lbl_xray_msu}, deemed conservative for this study.
A specific volume, modeled by an argon gas, is included in the plasma chamber to manage specifically the electrons bouncing back from the chamber walls toward the plasma. Indeed, for computation time reason, the source magnetic field is not included in the Fluka simulation and the straight trajectories of reflected electrons passing through the plasma chamber volume are consequently not realistic. Thus, an electron going back toward the plasma is considered as re-absorbed by the plasma and is stopped in the Fluka simulation. Because electrons are magnetized and approximately follow magnetic field lines in ECRIS, such a stopped electron is implicitly taken into account by starting a fresh new electron hitting the plasma chamber wall according to the MC output distribution. On the other hand, the secondary photons emitted backward toward the plasma can propagate freely through the gas volume, with a very small probability of interaction, like for crossing a plasma. Each of the three plasma chamber surface (injection disk, radial cylinder and extraction disk) that electrons hit is used to start an independent Fluka simulation. This allows in particular to investigate the secondary photon directionality. The 3 simulations results are next merged to obtain the full x-ray spectrum of the ion source.

\subsection{X-ray flux exiting the source and shielding scaling}

\begin{figure}[!ht]
\centering
\includegraphics[width=0.48\textwidth]{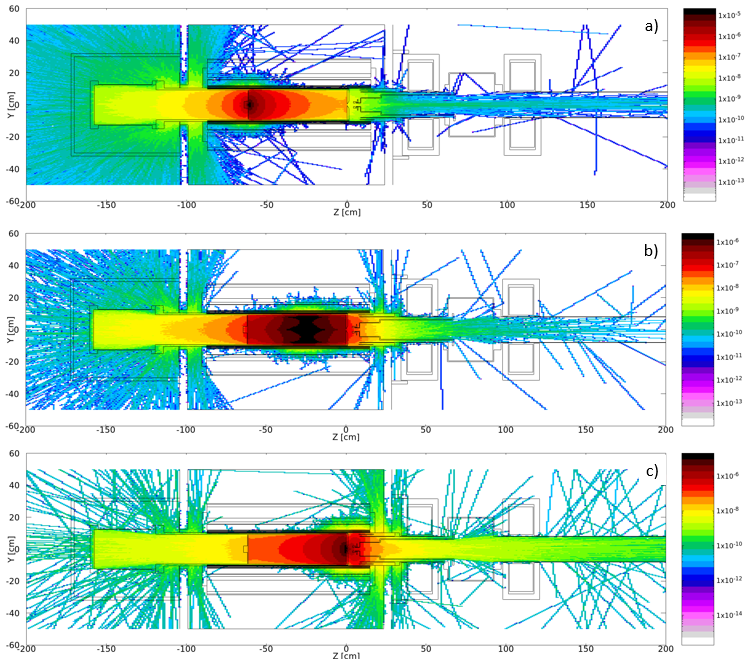}
\caption{Averaged x-ray fluence generated by the impact of a single electron on injection  (a), radial  (b), and extraction walls (c) when $B_{min}$\,=\,\SI{0.8}{\tesla} and $T_e$\,=\,120 keV.}
\label{fig:FIG6}
\end{figure}

\begin{figure}[!ht]
\centering
\includegraphics[width=0.48\textwidth]{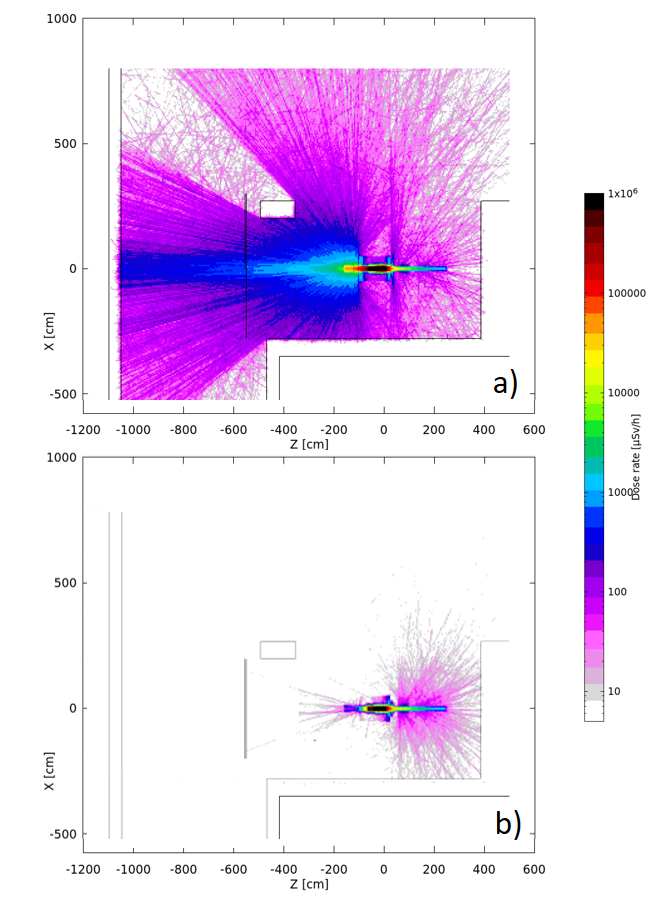}
\caption{X-ray dose per kW of electron  simulated in the cave (a) without and (b) with shielding when $B_{min}$\,=\,\SI{0.8}{\tesla} and $T_e$\,=\,\SI{120}{\kilo\electronvolt}.}
\label{fig:FIG7}
\end{figure}

\begin{figure}[!ht]
\centering
\includegraphics[width=0.48 \textwidth]{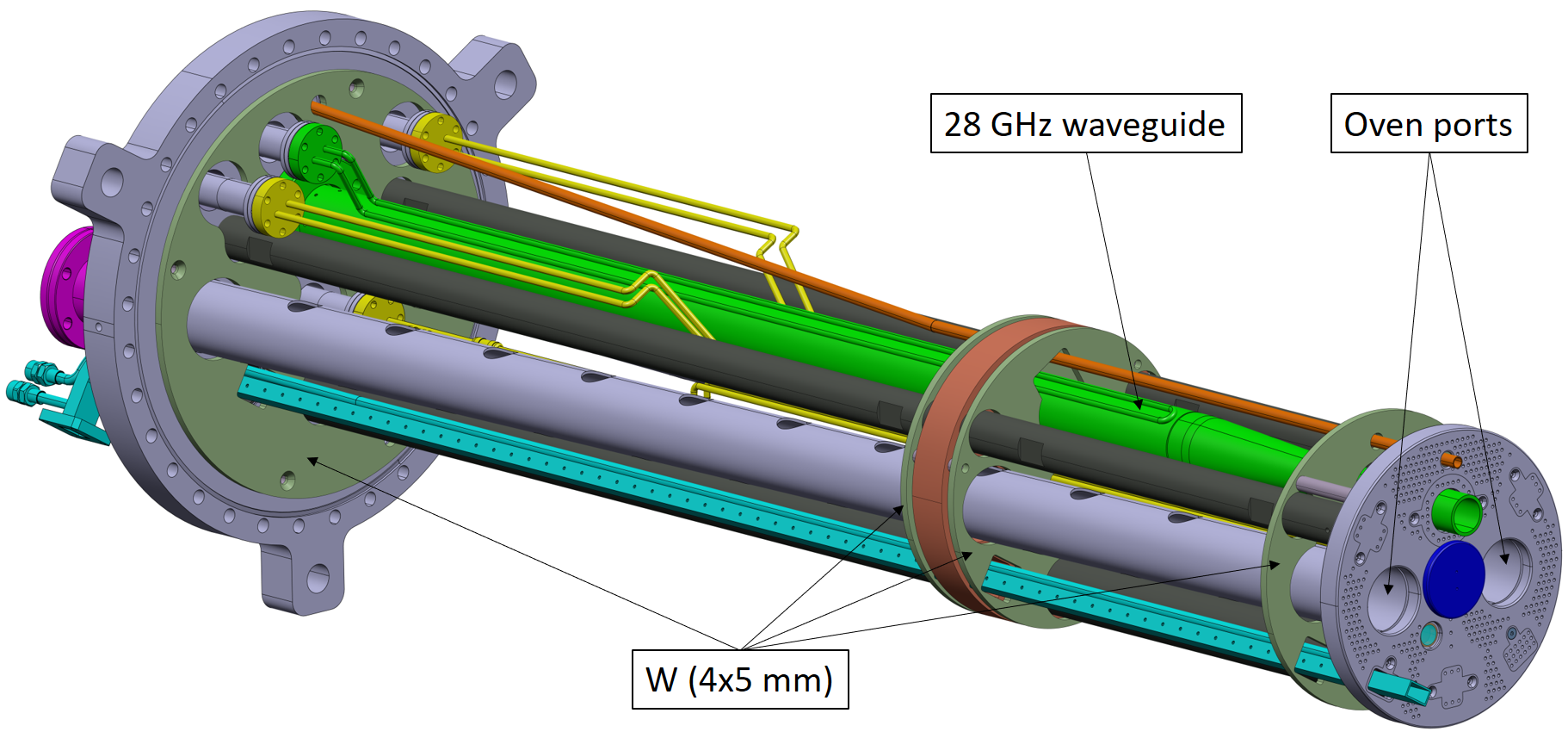}
\caption{View of the ion source injection system mechanics showing the 4 tungsten screens located under vacuum and the two specific ports modeled in Fluka: the two oven ports and the \SI{28}{\giga\hertz} over-sized waveguide. }
\label{fig:FIG8}
\end{figure}

In Fluka, the secondary particles showers (mainly photons and electrons) generated by each primary electron are followed until they are fully stopped by matter. Because of the small angle between the initial electron directions and the impacted surface, it is found that approximately 50\,\% of the electrons are bouncing back toward the plasma. Such a high reflection rate is likely to increase the actual electron confinement time in ECRIS. 
Figure~\ref{fig:FIG6} presents the  x-ray fluence per electron impinging (a) the injection surface, (b) the radial surface and (c) the extraction surface when $B_{min}$ and $T_e$ are 0.8~T and 120~keV respectively. One can see that the dominant x-ray photon leak able to exit the source occurs on the source injection side, at the place where the material thickness is the lowest. Being mainly composed of a thick layer of stainless steel, soft iron, Nb-Ti and copper, the superconducting magnet cold mass efficiently stops the radial photon flux in the first radial centimeters. The 5~cm thick iron yoke surrounding the superconducting magnet cryostat also strongly attenuates the x-ray flux passing through it. On the other hand, the x-rays exiting on the extraction side are either directed along the beam pipe axis or channeled radially in the gap between the source and the first focusing solenoid yoke. Table~\ref{tab:photonyield} presents the total x-ray yield per electron generated for $B_{min}$/$T_e$ =\,\SI{0.3}{\tesla}~/~\SI{50}{\kilo\electronvolt} and \SI{0.8}{\tesla}~/~\SI{120}{\kilo\electronvolt} respectively. The increase of $B_{min}$ from 0.3 to \SI{0.8}{\tesla} favors the leak of photons toward the injection and extraction walls (27\,\% and 62\,\% respectively) rather than the radial direction (11\,\% only). This effect is likely a consequence of the following facts: (i) an increase of the electron yield on the extraction wall for $B_{min}$\,=\,\SI{0.8}{\tesla} (see Tab.~\ref{tab:distr_position}), (ii) a high rate of electrons bouncing on the radial walls and (iii)  x-ray traversing thick material layers are subject to diffusion and back-scattering, favoring the remaining photons to escape toward the directions with lesser matter (injection and extraction). Indeed, Fluka propagates all the secondary particles generated by the incident electrons until the particle shower ends. The yields presented in Tab.~\ref{tab:photonyield} also include numerous lower energy photons subject to high scattering probability, which would finally not be detected outside of the ion source. 
 The dose around the source in the cave, when no specific shielding is present, is displayed in Fig~\ref{fig:FIG7}(a). The local dose is as high as \SI{100}{\micro\sievert\per\hour} per kW of injected electrons in the corridor located on the left (z\,<\,-600 cm) of the ion source, on its injection side. In order to damp the dose rate below the allowed 7.5 $\upmu$Sv limit in the corridor, specific shieldings were added and their effect investigated with  Fluka (see Fig.~\ref{fig:FIG5}). 
 In order to reduce the need of bulky shielding outside of the ion source on the injection side, a set of four 5 mm thick tungsten screens are placed under vacuum as close as possible from the x-ray emission, behind the plasma injection flange, as can be seen in Fig.~\ref{fig:FIG5} and ~\ref{fig:FIG8}. The shields  are hollowed by a set of three 40 mm diameter holes, to model the passage of the two metallic oven ports and the over-sized 28 GHz waveguide. The lack of matter inside these tubes is expected to channel x-rays out of the source and their effect had to be investigated. It is found that an extra layer of 20 mm of lead is required to shield the rear part of the source atom and RF injection system. This shield dimension is limited to a small solid angle and is located far away from the  place where the daily maintenance of the source is done.  Finally, a 5 mm thick lead shield is fixed along the fences closing the ion source high voltage zone (see the vertical line on the left of the source in Fig.~\ref{fig:FIG7}b). On the extraction side, the x-rays are blocked by a 5 mm thick tungsten cylinder located around the extraction system and by a 20 mm thick lead cylinder closing the radial gap between the source yoke and the extraction solenoid yoke. The resulting x-ray dose after filtering is displayed in Fig.~\ref{fig:FIG7}(b), showing only places where the dose is higher than \SI{5}{\micro\sievert\per\hour} per kW of electrons. One can check that the modeled compact shielding  efficiently screens the x-ray and prevent the dose to extend outside of the cave. The mean dose in the corridor on the left is lower than \SI{1}{\micro\sievert\per\hour}.
\begin{figure}[!ht]
\centering
\includegraphics[width=0.48\textwidth]{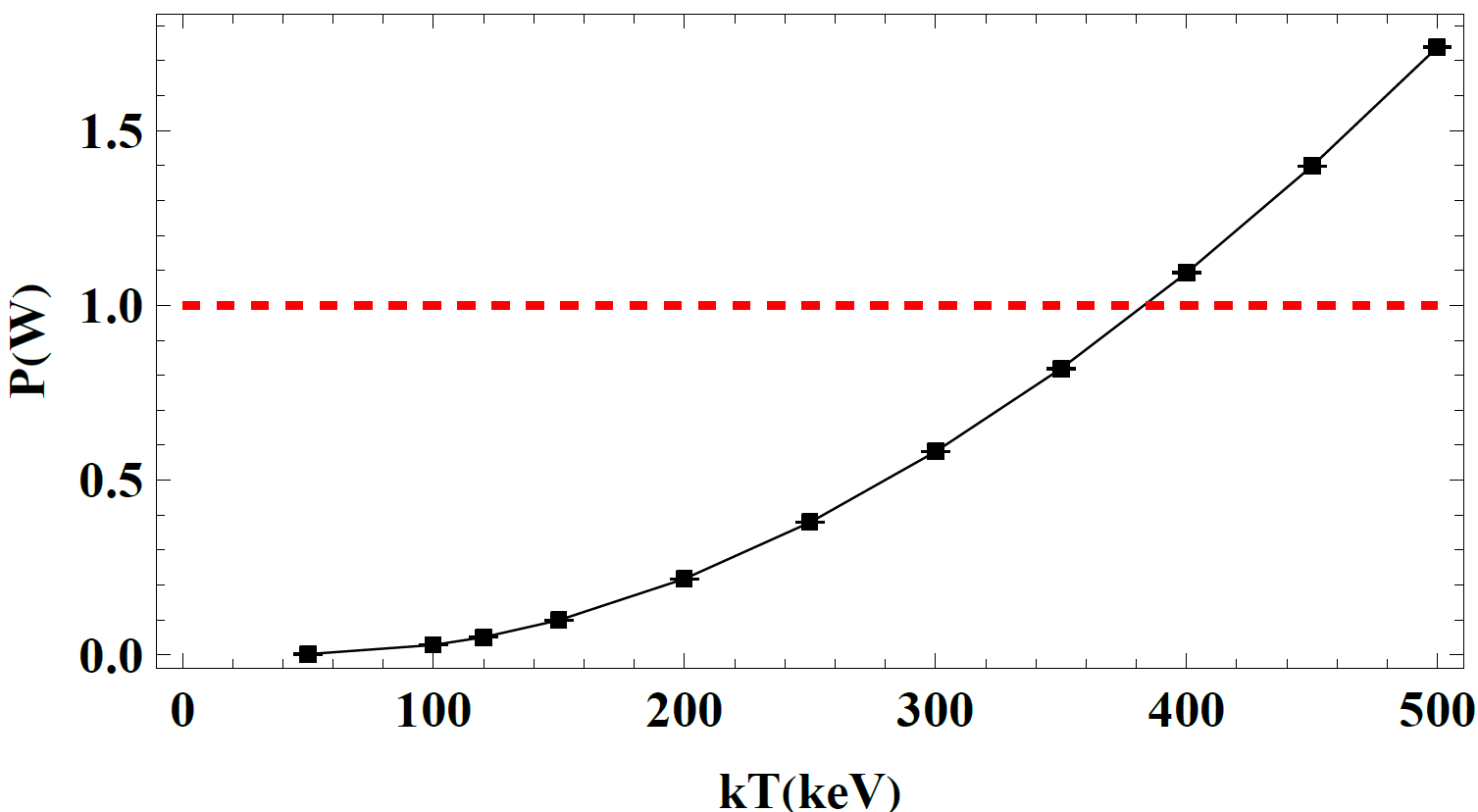}
\caption{Evolution of the total power deposited to the source cold mass (per kW of incident electrons) as a function of the EEDF temperature impinging the plasma chamber walls (black solid line). The 1 W level observed experimentally is indicated with a red dashed line.}
\label{fig:FIG9}
\end{figure}

\subsection{\label{sec:coldmass}X-ray power to the magnet cold mass }

The return of experience from the 3$^{rd}$ generation ECRIS operation at 28 GHz shows a parasitic heating of the superconducting magnet cold mass caused by the x-ray flux generated by electron bremsstrahlung impacting the plasma chamber walls. This heat load is linearly increasing with $B_{min}$ and values up to 1 W of extra heat load per kW of injected RF have been reported~\cite{bib:lbl_leitner_xray}. Fluka is used to investigate the power deposited into the cryostat as a function of the EEDF temperature. The electron initial direction corresponding to the case $B_{min}$\,=\,\SI{0.8}{\tesla} were used. The Maxwell-Boltzmann EEDF  temperature was varied from \SI{50}{\kilo\electronvolt} to \SI{500}{\kilo\electronvolt}. A set of $10^9$ electrons were cast in Fluka to obtain a sufficient statistic to investigate the power deposited in the cold mass per kW of impinging electrons to the plasma chamber wall. The resulting power is plotted in Fig.~\ref{fig:FIG9}. One can see that the EEDF temperature necessary to deposit 1 W per kW of electrons is $T~\approx$~\SI{380}{\kilo\electronvolt}. This value is 3 to 4 times higher than the x-ray spectral temperature experimentally measured on 3$^{rd}$ Generation ECRIS along the source axis~\cite{bib:lbl_leitner_xray,bib:lbl_xray_imp,bib:lbl_xray_msu}, suggesting an important temperature anisotropy in those ECRIS. This very high radial electron temperature obtained by Fluka is much higher than the one obtained with the MC code. A prospect to investigate this point is proposed in the conclusion. It is worth mentioning that high spectral temperature anisotropy has been experimentally reported with two distinct ECRIS having a (rare) direct radial view to the plasma. First with the Quadrumafios ion source, where the radial spectral temperature was measured up to \SI{230}{\kilo\electronvolt}, while the axial one was $\approx$~\SI{90}{\kilo\electronvolt} for a plasma heated at \SI{18}{\giga\hertz}~\cite{bib:gaudart,bib:Lamoureux}. And second, with the AECR source heated at 14+\SI{12.5}{\giga\hertz} which revealed a radial spectral temperature of \SI{100}{\kilo\electronvolt}, while the axial one was \SI{30}{\kilo\electronvolt}~\cite{bib:Noland}.

\section{Conclusion}
The x-ray dose exiting the ASTERICS ion source, induced by the impact of plasma electrons on the  plasma chamber wall, has been studied with Fluka, using as input the results of a MC electron code simulating the hot electron dynamics in the ECRIS. The MC code results indicate an important spatial electron temperature anisotropy at the three plasma chamber walls (injection disk, radial cylinder and extraction disk). An increase of the electron  temperature at the three walls is obtained when $B_{min}$ increases. The distribution of electron flux to these surfaces is also found to be strongly dependent on the value of $B_{min}$: the electron flux leaks preferentially toward the place where the magnetic field is minimum. This behavior is a consequence of the electron Coulomb scattering model used in the plasma. The distributions of angle of incidence of the electrons with respect to the normal to the local surface present a peak above 60° and 80° for the radial and extraction walls respectively. These distribution angles are a consequence of the magnetic mirroring effect happening in the strongly magnetized ECRIS
, which favors the confinement of electrons heated by the ECR mechanism and hence characterized by large perpendicular velocities. The large angle of incidence of electron to the wall results in a large amount of them ($\approx$\,50\,\%) being bounced back toward the plasma. It also results in specific solid angles of photon emissions that must be considered to appropriately simulate the x-ray spatial emission distribution from ECRIS. Without shielding, a dose higher than $\sim$\SI{100}{\micro\sievert\per\hour} per kW of primary electrons is obtained in the corridor located at a distance of 5~m from the injection side of the ECRIS in the NEWGAIN cave. A shielding composed of four \SI{5}{mm} plates of tungsten placed inside the source, under vacuum and as close as possible to the plasma chamber, is used to attenuate the x-ray dose emitted on the source injection side. This solution allows reducing dramatically the places where x-ray shielding must be placed around the ECRIS, resulting in a simplified ion source maintenance and a simpler and cheaper radiation safety design. A parametric EEDF temperature study done with Fluka finds that the electron temperature required to heat the ion source superconducting magnet cold mass up to \SI{1}{\watt} per kW of primary electrons, level of power measured experimentally, is \SI{380}{\kilo\electronvolt}. Such a high temperature is found neither with MC simulations nor by experimental photon spectral temperature axial measurements ($T_S\leq$\,\SI{100}{keV}), but this result is consistent with experimental spectral temperature anisotropy measured in two different ECRIS, where the electron radial spectral temperature was found to be 3-4 times higher than the axial one. A prospect to investigate this discrepancy is to implement a standing wave model in the MC code including the effect of the plasma, which will likely lead to an increase of the electric field seen by the electrons and an increase of the electron temperature of electrons heating the walls. Another important prospect of this work will be to map the actual x-ray dose emitted by the ion source and compare it with the present results.
\begin{acknowledgments}
This work is supported by Agence Nationale de la Recherche, with the contract \# 21-ESRE-0018 EQUIPEX+ NEWGAIN. The authors are grateful to Tanguy Cadoux (CEA-Irfu) for providing the ASTERICS cold mass geometry, to Mile Kusulja and Francis Vezzu (CNRS-LPSC) for providing the ion source mechanics dimensions and images and Guillaume Brunet (GANIL) for the LEBT geometry.
\end{acknowledgments}
\section*{Data Availability Statement}
The data that support the findings of this study are available from the corresponding author upon reasonable request. 
\nocite{*}
\bibliography{aipsamp}
\end{document}